\font\teneufm=eufm10 
\documentclass[12pt]{article}

\usepackage{amsfonts}
\begin{document}

\def\Eh{\mbox{\teneufm\char 83}}

\title{Hierarchy of Dirac, Pauli and Klein-Gordon conserved operators in  
Taub-NUT background}

\author{Ion I. Cot\u aescu \thanks{E-mail:~~~cota@physics.uvt.ro}\\ 
{\small \it West University of Timi\c soara,}\\
       {\small \it V. P\^ arvan Ave. 4, RO-1900 Timi\c soara, Romania}
\and
Mihai Visinescu \thanks{E-mail:~~~mvisin@theor1.theory.nipne.ro}\\
{\small \it Department of Theoretical Physics,}\\
{\small \it National Institute for Physics and Nuclear Engineering,}\\
{\small \it P.O.Box M.G.-6, Magurele, Bucharest, Romania}}
\date{\today}

\maketitle

\begin{abstract}
The algebra of conserved observables of the $SO(4,1)$ gauge-in\-va\-riant 
theory of the Dirac fermions in the external field of the Kaluza-Klein 
monopole is investigated. It is shown that the Dirac conserved operators 
have physical parts associated with Pauli operators that are also conserved 
in the sense of the Klein-Gordon theory. In this way one gets simpler 
methods of analyzing the properties of the conserved Dirac operators and 
their main algebraic structures including the representations of dynamical 
algebras governing the Dirac quantum modes.

Pacs 04.62.+v

\end{abstract}

\newpage

\section{Introduction}

The relativistic quantum mechanics, seen as the one-particle restriction of 
the Lagrangian quantum field theory on curved spacetimes, give rise to 
interesting mathematical problems concerning the properties of the physical  
observables. It is known that one of the largest algebras of conserved 
operators is produced by the Euclidean Taub-NUT geometry since beside usual 
isometries this has a hidden symmetry of the Kepler type \cite{GM,GRFH}. 
This is related to the existence of St\" ackel-Killing tensors connected 
with the components of an analogue to the Runge-Lenz vector of the Kepler 
type problem for which, in addition, can be expressed in terms of four 
Killing-Yano tensors \cite{GRFH,vH1,VV}.

The theory of the Dirac equation in the Kaluza-Klein monopole field was 
studied in the mid eighties \cite{DIRAC}. An attempt to take into account 
the Runge-Lenz vector of this problem was done in \cite{CH}. We have 
continued this study showing that the Dirac equation is analytically 
solvable \cite{CV2} and determining the energy eigenspinors of the 
central modes. Moreover, we derived all the conserved observables of 
this theory, including those associated with the hidden symmetries of 
the Taub-NUT geometry. Thus we obtained the Runge-Lenz vector-operator 
of the Dirac theory, pointing out its specific properties \cite{CV3}. 
The consequences of the existence of this  operator were studied 
in \cite{CV4} showing that the dynamical algebras of the Dirac theory 
corresponding to different spectral domains are the same as in the scalar 
case \cite{GRFH} but involving  other irreducible representations. Thus 
for the discrete energy spectra we  obtained two irreducible 
representations of the $o(4)$ algebra describing distinguish quantum 
modes for each energy level \cite{CV4}. 

This new phenomenon encourage us to continue the mathematical study 
of the whole algebra of conserved observables of the Dirac theory in 
Taub-NUT background. In our opinion, the  operators related to the 
manifest or hidden symmetries of the Taub-NUT geometry  are of a special 
interest since they reflect the effects of the geometry on the behavior 
of different quantum systems, with integer or half-integer spin. However, 
in the Dirac case there are several complicated operators whose  
manipulation can be  sometime  extremely difficult. We hope that a general 
study of their action on the Dirac spinors could lead to simpler 
calculation methods.    
     
The present article is devoted to this problem. Our goal is to separate the 
active parts of the conserved Dirac operators, called here physical parts, 
which determine the effects on the Dirac energy eigenspinors. Obviously 
these are projections obtained with the help of the projection operator on 
the whole space of physical spinors. We show that these are of a specific 
diagonal (even) or off-diagonal (odd) forms depending on Pauli operators 
that are also conserved in the sense of the Klein-Gordon theory. In this 
way we derive  simpler calculation rules and  identify associations among  
conserved Dirac and  Pauli operators as those currently used in theories 
involving monopoles \cite{DYON,JMP,HMH} or new other we write down here.

We start in the second section with a brief review of some previous results 
\cite{CV2,CV3,CV4} we need. In the next section we study the algebra of the 
conserved Dirac operators and introduce a new type of even operators which 
help us to define the projection operator that separates the physical parts. 
Furthermore, we point out that the diagonal (even) physical parts  of the 
Dirac observables can be associated with well-defined conserved Pauli 
operators obeying the same algebraic relations. In Section 4 we 
discuss the physical parts of the main conserved Dirac operators 
\cite{CV2,CV3,CV4} and we identify their associated Pauli conserved 
operators. Here after presenting the simplest conserved Pauli operators 
we derive those corresponding to isometries or hidden symmetries including 
the generators of the dynamical algebras. In this way we show that the 
results of \cite{CV4} hold also in the case 
of continuous energy spectra, for $so(3,1)$ or $e(3)$ dynamical algebras.
The conclusions and  comments  are presented in the last section and
in a short Appendix some formulas involving an important Pauli operator 
studied in \cite{JMP,HMH} are given.      
 
We work in natural units with $\hbar=c=1$.

\section{Preliminaries}

The background of the gauge-invariant five-dimensional theory of the Dirac 
fermions in the external field of the Kaluza-Klein monopole \cite{GPS} is 
the the Taub-NUT space with the time coordinate trivially added. It is 
convenient to consider the static chart of Cartesian coordinates $x^{\mu}$, 
($\mu, \nu,...=0,1,2,3,5$), with the line element
\begin{equation}\label{(met)} 
ds^{2}=g_{\mu\nu}dx^{\mu}dx^{\nu}=dt^{2}-\frac{1}{V}dl^{2}-V(dx^{5}+
A_{i}dx^{i})^{2}\,,
\end{equation}   
where $dl^{2}=(d\vec{x})^{2}=(dx^{1})^{2}+(dx^{2})^{2}+(dx^{3})^{2}$
is the usual Euclidean three-dimensional line element in  
Cartesian physical space coordinates $x^{i}$ ($i,j,...=1,2,3$). 
The other coordinates are the time, $x^{0}=t$, and 
the Cartesian Kaluza-Klein extra-coordinate, $x^{5}$. 
In (\ref{(met)}) the function $1/V(r)=1+\mu/r$ 
depends on $r=|\vec{x}|$ and the  
real parameter  $\mu$ while $A_{i}$ are the potentials of the Dirac 
monopole.  

This background has the isometry group  
$G_{s}=SO(3)\otimes U(1)_5\otimes T_{t}(1)$ formed by the rotations of the 
Cartesian space coordinates and $x^{5}$ and $t$ translations. The $U(1)_5$ 
symmetry is important since this eliminates the so called NUT singularity
if $x^5$ has the period $4\pi\mu$. The  Killing vectors $k_{(i)}$ 
($i=1,2,3$) and $k_{(5)}$  are directly connected with the conserved 
operators which appear in the scalar case. They can be expressed in terms 
of momentum operators 
$P_{i}=-i(\partial_{i}-A_{i}\partial_{5})$ and $P_{5}=-i\partial_{5}$
\cite{GRFH}. The last one, for negative mass models, can be interpreted 
as the ``relative electric charge"  and it is always conserved.
Moreover, the Taub-NUT geometry possesses four 
Killing-Yano tensors, $f^{(i)}_{\mu\nu}$ ($i=1,2,3$) and $f^Y_{\mu\nu}$, 
of valence 2, related to the hidden symmetries of the Taub-NUT geometry  
reflected by the existence of the non-trivial St\" ackel-Killing tensors 
$k_{(i)}^{\mu\nu}$ \cite{GRFH,VV,CV2,CV1}.

In this Kaluza-Klein geometry there is a pentad gauge fixing \cite{P} where 
the  massless Dirac field, 
$\psi$, satisfies a simple gauge-covariant Dirac equation, 
$\not{\!\!D}\psi=0$, where 
$\not{\!\!D}=i\gamma^{0}\partial_{t}-\not{\!\!D}_{s}$
 \cite{DKK,DIRAC,CV2,CV4}. 
In the standard representation of the Dirac matrices (with diagonal 
$\gamma^0$  \cite{TH}) the Hamiltonian operator \cite{CV2,CV4}
\begin{equation}\label{HH}
H =\gamma^0\not{\!\!D}_{s}
=\left(
\begin{array}{cc}
0&\alpha^{*}\\
\alpha&0
\end{array}\right)
\end{equation}
has manifest supersymmetry. It is expressed in terms of Pauli operators,  
\begin{eqnarray}
&\alpha&=\,\sqrt{V}\,\pi 
=\sqrt{V}\left( {\sigma}_{P}-\frac{iP_{5}}{V}\right)\,,\\ 
&\alpha^{*}&=\,V\pi^{*}\frac{1}{\sqrt{V}} 
=V\left({\sigma}_{P}+\frac{iP_{5}}{V}\right)\frac{1}{\sqrt{V}}\,, 
\end{eqnarray}
where $\sigma_P=\vec{\sigma}\cdot\vec{P}$ involves the Pauli matrices, 
$\sigma_i$. These  operators give the space part of the  Klein-Gordon 
operator as \cite{CV2,CV4},  
\begin{equation}
\Delta= \alpha^{*}\alpha=V\pi^{*}\pi= 
V{\vec{P}\,}^{2}+\frac{1}{V}{P_{5}}^{2}\,.
\end{equation}
We specify that the 
star superscript is a mere notation that does not represent the Hermitian 
conjugation because here we use a non-unitary representation of the 
algebra of Dirac operators. Of course, this is equivalent with the unitary 
representation where all of these operators are self-adjoint \cite{CV2}.

Since $P_5$ commutes with all the other conserved observables, we settle
its eigenvalue, $\hat q$,  such that $q\equiv -\mu\hat q=0,\pm 1/2,\pm 1,
...$ \cite{GM,GRFH}. 
We denote  by $\Eh$ the space of usual and generalized energy eigenspinors 
of the form  $U_{E}=(u_{E}, E^{-1}\alpha u_{E})^{T}$ which solve the 
eigenvalue problem $HU_{E}=EU_{E}$ \cite{CV2}. In \cite{CV2,CV4} we showed 
that $u_{E}$ is a solution of the static Klein-Gordon equation, 
$\Delta u_{E}=E^{2} u_{E}$,  that may be square integrable with respect to 
the specific relativistic scalar product of the Dirac theory \cite{CV2} or 
behave as tempered distributions. The Klein-Gordon equation is analytically 
solvable producing continuous energy spectra,  $E\ge |\hat q|$, 
for any real $\mu$ and discrete energy levels, $E_{n}$ with $n>|q|>0$, 
only for $\mu<0$. These are included in the domain $(0,\hat q)$ such that 
$\lim_{n\to \infty}E_{n}=|\hat q|$. Hence,  there are no zero modes 
and the operator $\Delta$ can be considered invertible on the space of  
of the spinors $u_{E}$.    
Therefore, we can conclude that the Dirac equation produces the {\em same} 
energy spectra as the Klein-Gordon one. Moreover, since there are no zero 
modes the Hamiltonian operator (\ref{HH}) is also invertible. The meaning 
of this operation will be discussed in the next section.

\section{The algebra of conserved Dirac operators}   

Our aim is to study here  the form and the action of the {\em conserved} 
operators of the Dirac theory which, by definition, are the Dirac operators 
that {\em commute} with the Hamiltonian (\ref{HH}). We denote by  ${\bf D}$ 
the algebra of these operators. We say that a Pauli operator $\hat X$ acting 
on the space of the two-component Pauli spinors $u_{E}$ is {\em conserved} 
if it commutes with $\Delta$. We denote by ${\bf P}$ the algebra of the 
conserved Pauli operators, including the conserved observables of the  
Klein-Gordon theory, called here {\em orbital} operators. In this section we 
denote systematically by capitals the operators of ${\bf D}$ and by hated ones 
the operators of ${\bf P}$ without to use special notations for the identity 
operators of these algebras.

In general, the Pauli blocks, $\hat X^{(ab)}$  ($a,b =1,2$), of any 
conserved Dirac operator
\begin{equation}\label{Xab}
X=\left(
\begin{array}{cc}
\hat X^{(11)}&\hat X^{(12)}\\
\hat X^{(21)}&\hat X^{(22)}
\end{array}\right)\,\in {\bf D}
\end{equation}
satisfy the conditions  
\begin{eqnarray}
&&\hat X^{(22)}\alpha= \alpha \hat X^{(11)}\,,\quad
\alpha^{*}\hat X^{(22)}=
\hat X^{(11)}\alpha^{*}\label{X1}\,, \\
&&\hat X^{(12)}\alpha= \alpha^{*} \hat X^{(21)}
\,,\quad 
\alpha \hat X^{(12)}=
\hat X^{(21)}\alpha^{*}\label{X2}
\end{eqnarray}
which are equivalent with $[X,H]=0$. Hereby it results that 
\begin{equation}
\hat X^{(21)}=\alpha \hat X^{(12)}\alpha \Delta^{-1}
\end{equation}
 and
\begin{equation}
[\hat X^{(11)},\Delta]=
[\hat X^{(12)}\alpha,\Delta]=
[\alpha^{*}\hat X^{(21)},\Delta]=0
\end{equation} 
which means that $\hat X^{(11)},\,\hat X^{(12)}\alpha,\,
\alpha^{*}\hat X^{(21)}\in {\bf P}$.

We observe that  possible solutions of Eqs.(\ref{X1}) and (\ref{X2}) are 
the diagonal operators 
\begin{equation}
{\cal D}(\hat X)=\left(
\begin{array}{cc}
\hat X&0\\
0&\alpha\hat X\Delta^{-1}\alpha^{*}
\end{array}\right)
\end{equation}
where $\hat X\in {\bf P}$. Particularly, for $\hat X=1$ we obtain the 
projection operator 
\begin{equation}\label{id} 
I ={\cal D}(1)=\left(
\begin{array}{cc}
1&0\\
0&\alpha\Delta^{-1}\alpha^{*}
\end{array}\right)
\end{equation}
on the space $\Eh$. This split the algebra 
${\bf D}={\bf D}_{0}\oplus {\bf D}_{1}$ in two subspaces of the projections 
$XI\in {\bf D}_{0}$ and $X(1-I)\in {\bf D}_{1}$ of all $X\in {\bf D}$. 
According to Eqs.(\ref{X1}) and (\ref{X2}) we find that the projections of 
two arbitrary operators $X,\,Y\in {\bf D}$ satisfy  
$(XI)(YI)=(XY)I$ and $[X(1-I)](YI)=0$ which lead to the conclusion that    
${\bf D}_{0}$ is a subalgebra while ${\bf D}_{1}$ is even an ideal in  
${\bf D}$. Obviously, $I$ is the identity operator of ${\bf D}_{0}$. 
On the other hand, in \cite{CV2} we introduced  the ${\cal Q}$-operators 
defined as 
\begin{equation}
{\cal Q}(\hat X)=\left\{ H\,,\,\left(
\begin{array}{cc}
\hat X&0\\
0&0
\end{array} 
\right) \right\}
=\left(
\begin{array}{cc}
0&\hat X\alpha^*\\
\alpha\hat X&0
\end{array} 
\right)\,, 
\end{equation}
where $\hat X$ may be any Pauli operator. However, if $\hat 
X\in {\bf P}$ then ${\cal Q}(\hat X)\in {\bf D}_{0}$ since  
$[{\cal Q}(\hat X),H]=0$ and ${\cal Q}(\hat X)I= {\cal Q}(\hat X)$. 
If $\hat X=1$ we obtain just the Hamiltonian operator 
$H={\cal Q}(1)\in {\bf D}_0$. Consequently, the inverse of $H$ with respect 
to $I$ can be represented as $H^{-1}={\cal Q}(\Delta^{-1})$. 
The mappings ${\cal D} : {\bf P}\to {\bf D}_{0}$ and
${\cal Q} : {\bf P}\to {\bf D}_{0}$ are linear and have the following 
algebraic properties
\begin{eqnarray}
{\cal D}(\hat X){\cal D}(\hat Y)&=&{\cal D}(\hat X \hat Y)\,,\\
{\cal Q}(\hat X){\cal Q}(\hat Y)&=&{\cal D}(\hat X \hat Y\Delta)\,,\\
{\cal D}(\hat X){\cal Q}(\hat Y)&=&
{\cal Q}(\hat X){\cal D}(\hat Y)={\cal Q}(\hat X \hat Y)\,,
\end{eqnarray}
for any  $\hat X,\, \hat Y \in {\bf P}$. Moreover, the relations    
\begin{equation}
[\gamma^0,\, {\cal D}(\hat X)]=0\,, \quad
\{\gamma^0,\, {\cal Q}(\hat X)\}=0 
\end{equation}
indicate that, according to the usual terminology \cite{TH},   
${\cal D}$ and $\gamma^0{\cal D}$ are {\em even} Dirac operators while 
${\cal Q}$ and $\gamma^0{\cal Q}$ are {\em odd} ones. We note that there 
are many other odd or even operators which do not have such forms.

In general, since $I$ is the projection operator on the space of the Dirac 
energy eigenspinors $\Eh$, we say that the projection $IXI$ of any Dirac 
operator $X$, conserved or not, represents the {\em physical part} of $X$. 
We can convince ourselves that if $X\in {\bf D}$ then
\begin{equation}\label{WW}
IXI\equiv XI={\cal D}(\hat X^{(11)})+{\cal Q}(\hat X^{(12)}\alpha 
\Delta^{-1})\,,
\end{equation}
which means that all the operators from ${\bf D}_{0}$ can be written as 
${\cal D}$ or ${\cal Q}$-operators. Thus the action of $X$ reduces to 
that of the Pauli operators involved in (\ref{WW}) allowing us to rewrite 
the problems of the Dirac theory in terms  of Pauli operators 
\cite{CV3,CV4}. Indeed, it is easy to show that the action of any  operator 
$X\in {\bf D}$ on $U_{E}\in \Eh$  is  
\begin{equation}\label{cue}
 XU_{E}=XIU_{E}=\left(
\begin{array}{r}
\hat{\cal P}_{E}(X)\,u_{E}\\
E^{-1}\alpha\hat{\cal P}_{E}(X)\,u_{E}
\end{array}\right)\,,
\end{equation}
where, by definition, 
\begin{equation}\label{PE}
\hat{\cal P}_{E}(X)=\hat X^{(11)}+ E^{-1}\hat X^{(12)}\alpha
\end{equation}
is the {\em conserved} Pauli operator {\em associated} to $X$. 
Since the mapping $\hat{\cal P}_{E}: {\bf D}\to {\bf P}$ is linear and 
satisfies  $\hat{\cal P}_{E}(X)=  \hat{\cal P}_{E}(XI)$  it results that 
${\rm Ker}\,\hat{\cal P}_{E}={\bf D}_{1}$. 
In other respects,  Eqs.(\ref{X1}) and (\ref{X2}) lead to 
the  important property
\begin{equation}
\hat{\cal P}_{E}(XY)=  \hat{\cal P}_{E}(X)\hat{\cal P}_{E}(Y)\,, \quad \forall 
X,Y\in {\bf D}\,.
\end{equation}
which guarantees  that  $\hat{\cal P}_{E}$ preserves  the algebraic 
relations, mapping any algebra or superalgebra of ${\bf D}_0$ into an 
{\em isomorphic} algebra or superalgebra of ${\bf P}$, with the same 
commutation and anticommutation rules.

\section{Conserved observables}

In what follows we focus on the physical parts of the main Dirac 
conserved observables  pointing out the technical advantages 
of using ${\cal D}$ and ${\cal Q}$-operators that help us to identify the 
associated Pauli operators defined by (\ref{PE}). The even physical 
parts, ${\cal D}$, are associated to Pauli operators independent on $E$  
which are, therefore, well-defined  physical observables. For this reason it 
is useful to briefly review the most important conserved Pauli operators and 
then turn to the physical parts of the Dirac  ones. 

\subsection{Conserved orbital and Pauli operators}

In general, the Pauli operators are $2\times 2$ matrix differential 
operators acting on two-component Pauli spinors. There are many 
non conserved operators which do not commute with $\Delta$ as, for example, 
$\alpha,\,
\alpha^{*},\,\pi,\,\pi^{*},\, \sigma_{r}=\vec{\sigma}\cdot\vec{x}/r$ or
the operator $\lambda=\vec{\sigma}\cdot(\vec{x}\times \vec{P})+1$ proposed 
in \cite{JMP} and discussed in \cite{HMH}. Some algebraic properties of 
these operators are given in Appendix. 
 
By definition, the conserved operators of ${\bf P}$ commute with $\Delta$ 
which is the static part of the Klein-Gordon operator. As mentioned, these 
can be the usual  conserved  orbital operators of the scalar fields or more 
complicated ones involving, in addition, the  Pauli matrices which 
also commute with $\Delta$.     

The main conserved orbital operators are  the basis generators of the 
natural representation of the group $G_{s}$ carried by the space of 
scalar fields. These generators are defined up to the factor 
$-i$ as the Killing vector fields corresponding to isometries or the 
operators given by the Killing tensors associated to the hidden symmetries. 
The  $U_{5}(1)$ generator is $P_{5}$ and the  Killing vectors 
$k_{(i)}^{\mu}$ give the $SO(3)$ generators which are the components $L_i$ 
of the orbital angular momentum operator \cite{CV2,CV4}
\begin{equation}\label{(angmom)}
\vec{L}\,=\,\vec{x}\times\vec{P}-\mu\frac{\vec{x}}{r}P_{5}\,.
\end{equation} 
These commute with $\Delta$ and satisfy the canonical commutation 
relations among themselves and with the components of all the other 
vector operators (e.g. coordinates,  momenta, etc.). On the other hand, 
the specific Killing tensors $k_{(i)}^{\mu\nu}$ of the Taub-NUT geometry 
allow one to define the Runge-Lenz operator for scalar particles 
\cite{GRFH}
\begin{equation}\label{RLorb}   
\vec{K}=
\frac{1}{2}(\vec{P}\times \vec{L}-\vec{L}\times \vec{P})-
\frac{\mu}{2}\frac{\vec{x}}{r}\Delta +\mu\frac{\vec{x}}{r}{P_{5}}^{2}\,,
\end{equation}
which commute with $\Delta$ and its components satisfy the commutation 
relations
\begin{equation}\label{algKL1}
\left[ L_{i},\, K_{j} \right] = i \varepsilon_{ijk}\,K_{k}\,,\quad
\left[ K_{i},\, K_{j} \right] = i \varepsilon_{ijk}L_{k}F^2\,, 
\end{equation}
where $F^2 ={P_5}^2-\Delta$. For given values of $E$ and $\hat q$ this 
operator can be re-scaled in order to  recover the dynamical algebras 
corresponding to different spectral domains of the Kepler-type problems  
\cite{GRFH}. The new operators
\begin{equation}\label{Ri}
R_{i}=\left\{  
\begin{array}{lllll}
{ F}^{-1}{ K}_{i}&{\rm for}& \mu<0&{\rm and}&E<|\hat q|\\
{ K}_{i}&{\rm for}&{\rm any}~ \mu&{\rm and}&E=|\hat q|\\
\pm i{ F}^{-1}{ K}_{i}&{\rm for}&{\rm any}~ \mu &{\rm and}& E>|\hat q|
\end{array}\right.
\end{equation}
and $L_i$ ($i=1,2,3$)  generate either a  representation of the $o(4)$ 
algebra for the discrete energy spectrum in the domain $0<E<|\hat q|$ or 
a representation of the $o(3,1)$ algebra for  continuous spectrum in the 
domain $E>|\hat q|$. A special case is that of the  dynamical algebra  
$e(3)$ which  corresponds only to the ground energy of the continuous 
spectrum, $E=|\hat q|$. 

The operators of ${\bf P}$ involving Pauli matrices can be vector 
operators as the total angular momentum, 
\begin{equation}\label{JP}
\vec{J} =\vec{L}+\frac{\vec{\sigma}}{2}\,,
\end{equation}
or scalar operators of the form $\sigma_L =\vec{\sigma}\cdot\vec{L}$,\,
$\sigma_K =\vec{\sigma}\cdot\vec{K}$ or  
$\sigma_R =\vec{\sigma}\cdot\vec{R}$,
involved in superalgebras as
\begin{equation}\label{sksl} 
\left\{\sigma_{K},\,\sigma_{L}+1\right\}=0\,.
\end{equation}
Other conserved Pauli operators with more complicated structure have 
to be derived in association with the physical parts of the conserved 
Dirac observables. 
 
\subsection{Associated Dirac and Pauli operators}

We have seen that the physical parts of the conserved Dirac observables 
can have diagonal or off-diagonal terms among them only the diagonal 
ones can be correctly associated to  conserved Pauli operators independent 
on $E$.  However, the off-diagonal operators can be transformed at any 
time in diagonal ones using the multiplication with $H$ or $H^{-1}$. 
For example,  $H$ itself which is off-diagonal is related to the diagonal 
operators $H^2={\cal D}(\Delta)$ or  $I$. Thus each conserved Dirac 
operator can be brought in a diagonal form  associated with an operator 
from ${\bf P}$. 

Let us start with the generators of the representations of the group 
$G_{s}$ carried by the space of the Dirac spinors. The $U(1)_{5}$ 
generator remains the former  operator $P_{5}$ but the $SO(3)$ generators 
get  the usual spin terms, 
$S_{i}=\frac{1}{2}{\rm diag}(\sigma_{i},\,\sigma_{i})$ of the total 
angular momentum whose components, ${\cal J}_{i}=L_{i} +S_{i}$, 
commute with $H$ even if neither $L_{i}$ nor $S_{i}$ do not have this 
property \cite{CV2,CV4}. However, the effect on the spinors of $\Eh$ 
is due only to the physical parts which read 
\begin{equation}\label{JI}
{\cal J}_{i}I= {\cal D}(J_{i})={\cal D}(L_{i}) + \frac{1}{2}
{\cal D}(\sigma_{i})
\end{equation}
where both the orbital and the spin terms are {\em separately} conserved 
since $L_{i}$ and $\sigma_{i}$ commute with $\Delta$. Obviously, in this 
case the associated Pauli operators are just $J_i$ defined by (\ref{JP}).

The simplest conserved off-diagonal operators  are the so called 
Dirac-type operators  generated by the first three Killing-Yano tensors, 
$f^{(i)}$. We have shown  \cite{CV2,CV3} that these can be written simply 
in the form $Q_{i}={\cal Q}(\sigma_{i})$ which explains why their 
algebraic properties are close to those of the Pauli matrices. Now, we 
can prove that the diagonal operators $H^{-1}Q_{i}={\cal D}(\sigma_{i})$  
form a representation of the algebra of Pauli matrices with values in 
${\bf D}_{0}$ since
\begin{equation}
H^{-1}Q_{i}\,H^{-1}Q_{j}=\delta_{ij}I+i\varepsilon_{ijk}H^{-1}Q_{k}\,.
\end{equation}

The corresponding Dirac-type operator of the last Killing-Yano tensor, 
$f^Y$, calculated  according to the general rule of \cite{CML} has been 
obtained in \cite{CV3}.  This has the form
\begin{equation}\label{dy1}
Q^Y=-{\cal Q}(\sigma_r)
+\frac{2i}{\mu\sqrt{V}}\left(
\begin{array}{cc}
0&\lambda\\
-\lambda&0
\end{array}
\right)\,.
\end{equation}
Using the identities presented in Appendix one finds the equivalent forms 
reported in \cite{CV3}  and verify that $Q^Y$ commutes with $H$ and 
$P_{5}$ and anticommutes with $\not{\!\!D}_s$ and $\gamma^0$. Moreover, 
after a little calculation,  we obtain the remarkable identity
\begin{equation}
\mu P_{5}\left[Q^{Y}+ {\cal Q}(\sigma_r)\right]=\left\{H,\,\Lambda\right\}
\end{equation}
involving the operator $\Lambda={\rm diag}(\lambda,\lambda)$ that is a 
particular version of the Biedenharn operator  \cite{BID}. This is not 
conserved but 
$\Lambda^{2}=\vec{{\cal J}}^2-\mu^{2}{P_{5}}^{2}+1/4$ has this property.
Furthermore, we observe that, according to (\ref{Alam}) and 
(\ref{silam}), the physical part of $Q^Y$  can be put in the form 
\begin{equation}
Q^{Y}I={\cal Q}\left(-\sigma_{r}+\frac{2i}{\mu}\lambda\pi\Delta^{-1}\right)
={\cal Q}(\sigma^{Y}\Delta^{-1}) \,,
\end{equation}
where
\begin{equation}
\sigma^{Y}=\frac{2}{\mu}\left[\sigma_{K}+(\sigma_{L}+1)P_{5}\right]
\end{equation} 
is a new conserved Pauli operator  associated to $H Q^{Y}=H Q^{Y}I= 
{\cal D}(\sigma^{Y})$.   

\subsection{The Runge-Lenz operator and dynamical algebras}

These results allow us to calculate directly the physical parts of the 
Runge-Lenz operator of the Dirac theory (related to the Killing tensor 
$\vec{k}^{\mu\nu}$), following the same procedure as in \cite{CV3}.  
We start with the equivalent definition of the physical parts of the 
auxiliary operators \cite{CV3}   
\begin{equation}
{\cal N}_{i}I=\frac{\mu}{4}\left\{HQ^{Y}\,,\,H^{-1}Q_{i}\right\}-
{\cal J}_{i}P_{5}I\,,
\end{equation}   
which can be written as 
\begin{equation}
{\cal N}_{i}I={\cal D}\left(\frac{\mu}{4}\left\{ \sigma^{Y}\,,\,\sigma_{i}
\right\}-
 J_{i}P_{5}\right)={\cal D}(\hat N_{i})\,,
\end{equation}   
in terms of their associated conserved Pauli operators, 
\begin{equation}\label{Nas}
\hat N_{i}=K_{i}+\frac{\sigma_{i}}{2}\,P_{5}\,.
\end{equation}
Furthermore, we define the physical parts of the components of the 
conserved Runge-Lenz operator \cite{CV3,CV4} 
\begin{equation}\label{KI}
{\cal K}_{i}I={\cal N}_{i}I+ \frac{1}{2}({\cal F}-P_5)H^{-1} Q_i
\end{equation}
where ${\cal F}^{2}={P_{5}}^{2}-H^2$. Since  ${\cal F}^2 I={\cal D}(F^2)$,  
we can express  ${\cal K}_{i}I= {\cal D}(\hat K_{i})$, now the    
associated conserved Pauli operators being, 
\begin{equation}\label{Kas}
\hat K_{i}=K_{i} +\frac{\sigma_{i}}{2}\,F\,.
\end{equation} 
All these associations help us to understand the significance 
of the isomorphism among the algebra of the Dirac operators 
\cite{CV3,CV4}, 
\begin{equation}
\left[ {\cal J}_{i},\, {\cal K}_{j}\right]=i\varepsilon _{ijk}{\cal K}_{k}
\,,\quad
\left[ {\cal K}_{i},\, {\cal K}_{j}\right]=i\varepsilon _{ijk}{\cal J}_{k}
{\cal F}^{2}\,,
\end{equation}
that of the Pauli operators,
\begin{equation}
\left[  J_{i},\, \hat K_{j}\right]=i\varepsilon _{ijk}{\hat K}_{k}
\,, \quad
\left[ {\hat K}_{i},\, {\hat K}_{j}\right]=i\varepsilon _{ijk} J_{k}
 F^{2}\,,
\end{equation}
and (\ref{algKL1}).

Re-scaling (\ref{KI}) as in the case of the orbital operators (\ref{Ri}), 
but using ${\cal F}$ instead of $F$, one obtains the even operators 
${\cal R}_{i}\in {\bf D}$ \cite{CV4} having simple physical parts, 
${\cal R}_{i}I={\cal D}(\hat R_i)$, associated with the conserved 
Pauli operators 
\begin{equation}
\hat R_{i}=\left\{
\begin{array}{lcl}
R_{i}+\frac{\textstyle \sigma_i}{\textstyle2}&{\rm for}&E \not=\hat q\\ 
K_{i}&{\rm for}&E =\hat q 
\end{array}\right. \,.
\end{equation}
We specify that the orbital and  spin terms of 
${\cal N}_{i}I$, ${\cal K}_i I$ 
and ${\cal R}_{i}I$ (for $E\not =\hat q$) are also separately conserved, 
as in the case of the angular momentum, since $K_i$ and $F$ commute with 
$\Delta$.

The representations of the dynamical algebras $o(4)$ or $o(3,1)$ that 
govern the Dirac modes for $E\not=\hat q$ are generated by ${\cal J}_{i}$ 
and ${\cal R}_{i}$ \cite{CV4} whose physical parts, ${\cal J}_{i}I$ and 
${\cal R}_{i}I$, have the same spin terms, ${\cal D}(\sigma_{i})/2$. 
Therefore,  each of these representations is the direct product between 
the irreducible representation of scalar modes and a spin half 
two-dimensional (fundamental) representation of the dynamical algebra 
\cite{CV4}. When $E=\hat q$ then ${\cal F}$ and $F$ vanish  such that 
the representation of the subalgebra $so(3)\subset e(3)$ remains 
generated by the operators (\ref{JI}) while the operators  
${\cal K}_{i}I$ lose their spin terms becoming the translation 
generators of $e(3)$. All these representations arising from direct 
products are reducible. We note that this phenomenon is new since in 
the scalar (Klein-Gordon) case the representations of the dynamical 
algebras of the Kepler-type problems are irreducible \cite{GRFH}.  
However, these results could be easily obtained analyzing the 
{\em equivalent} representations generated by the associated Pauli 
operators $J_{i}$ and $\hat R_{i}$  as we did already in \cite{CV4} for 
the discrete energy spectrum.   We recall that therein we introduced 
the new conserved operator ${\cal C}=2\vec{\cal J}\cdot\vec{\cal R}-1/2$  
in order to distinguish between the irreducible representations of the 
$o(4)$ dynamical algebra. Now we see that 
${\cal C}I={\cal D}(\sigma_{R}+\sigma_{L}+1)$ where, according to 
(\ref{sksl}), we have $\{\sigma_{R},\,\sigma_{L}+1\}=0$. 

\section{Conclusions}

The first conclusion is that our approach allows one to associate the 
conserved Dirac operators of diagonal (even) form to conserved Pauli 
operators independent on $E$. Thus for each type of symmetry we have 
conserved operators at three levels: Dirac, Pauli and orbital (of the 
Klein-Gordon theory). 
The following table resumes this hierarchy (K is an abbreviation for 
Killing, K-Y for Killing-Yano; * denotes entries which involve issues 
too complex to be abbreviated in the table and some comments are given 
below).

~

\begin{center}
\begin{tabular}{cccccc}
\hline
geometric       &nature   &symmetry&Dirac   &Pauli   &Klein-Gordon\\
object          &         &        &operator&operator&operator\\      
&&&&&\\
\hline
&&&&&\\
$f_{\mu\nu}^{(i)}$&K-Y tensor&*     &$H^{-1}Q_{i}$&$\sigma_{i}$&-\\
$f_{\mu\nu}^{Y}$&K-Y tensor&*      &$HQ^{Y}$&$\sigma^{Y}$&-\\
$k_{(5)}^{\mu}$&K vector&$U(1)_5$&$P_{5}$&$P_{5}$&$P_{5}$\\
$k_{(i)}^{\mu}$&K vector&$SO(3)$&${\cal J}_{i}$&$J_{i}$&$L_{i}$\\
$k_{(i)}^{\mu\nu}$&K tensor& hidden&${\cal K}_{i},{\cal R}_i$
&$\hat{K}_{i},\hat{R}_{i} $&$K_{i},R_i$\\
\hline
\end{tabular}
\end{center}

~

However, there are many other even or odd conserved Dirac operators (e.g., 
${\cal D}(\sigma_{K}^2)$, ${\cal Q}(L_{i})$, ${\cal Q}(\sigma^Y)$, etc.) 
which can be constructed with the help of the conserved Pauli or orbital 
ones. This large collection of conserved observables is in fact a rich 
algebra freely generated by those related to the manifest or hidden 
symmetries of the Taub-NUT geometry.  

In $N=1$ supersymmetric quantum models with standard supersymmetry 
there is a single supercharge $Q$ that closes $Q^2 = H$  on the 
Hamiltonian. In many of these models, and that is the case of the 
Taub-NUT manifold, one can find additional or hidden, non standard 
supercharges involving Killing-Yano tensors. 
The Killing-Yano tensors $f^{(i)}_{\mu\nu}$ ($i=1,2,3$) give a vector 
representation of $SO(3)$ and their existence is connected with the 
complex structures of the hyper-K\"ahler Taub-NUT space. 
The forth Killing-Yano tensor $f^Y_{\mu\nu}$ is a singlet and exists by 
virtue of the metric being type $D$. All four Killing-Yano tensors are 
invariant under the action of $U(1)_5$ which physically represents the 
relative electric charge of two monopoles.

For spin-${1\over 2}$ particles, the Killing-Yano tensors are essential 
in construction of Dirac-type operators and evaluation of the spin 
contributions to the conserved quantities from the scalar case.  The 
antisymmetric feature of these operators make them the natural object 
used in description of the Dirac fermion in a curved spacetime. On the 
other hand, the fact that the St\" ackel-Killing tensors involved in 
the Runge-Lenz vector (\ref{RLorb}) can be expressed as symmetrized 
products of Killing-Yano tensors seems to be useless for scalar 
particles described by Schr\" odinger or Klein-Gordon equations. 
Therefore the existence of a certain square root of the 
St\" ackel-Killing tensors becomes relevant only in the presence of 
fermions.

In other respects, we can eliminate many difficulties due to the spin terms 
of the Dirac theory if we restrict ourselves only to 
the physical parts $XI$ of the operators $X\in {\bf D}$ giving up the 
projections $X(1-I)\in {\bf D}_{1}$  which can give rise  sometime to 
very complicated calculations. Moreover, we get the advantage  
of reducing the algebraic operations among the physical parts from 
${\bf D}_0$ to calculations involving only the associated Pauli operators 
from ${\bf P}$. For example, if instead of $[\vec{Q},\,Q^Y]$, we calculate 
only its physical part,  $[\vec{Q},\,Q^Y]I=[H^{-1}\vec{Q},\,HQ^Y]= 
{\cal D}([\vec{\sigma},\,\sigma^Y])$, we avoid a tedious algebra
easily obtaining the interesting identity
\begin{equation}
\frac{\mu}{4}\,[\vec{Q},\,Q^YI]= i(\vec{\cal K}+\vec{\cal J}P_{5})\times 
(H^{-1}\vec{Q})  + (F+P_5)\,H^{-1}\vec{Q}
\end{equation}  
which shows that this commutator does not produce new conserved 
observables.

Finally we note that our method based on the separation of the 
physical parts expressed in terms of ${\cal D}$ and ${\cal Q}$-operators 
could be used in any problem where the Hamiltonian is invertible and has 
manifest supersymmetry.

\setcounter{equation}{0} \renewcommand{\theequation}
{A.\arabic{equation}}

\section*{Appendix A: The operator $\lambda$} 

The operator
\begin{equation}\label{Alam}
\lambda=\vec{\sigma}\cdot(\vec{x}\times \vec{P})+1
=\sigma_{L}+1+\mu\sigma_{r}P_{5}
\end{equation} 
has the properties
\begin{equation}
\{\sigma_{r},\lambda\}=0\,,\quad [\sigma_{r},\sigma_{P}]=
\frac{2i}{r}\,\lambda
\end{equation}
and
\begin{equation}\label{silam}
\sigma_{P}\lambda=-\lambda\sigma_{P}=
\frac{1}{2}\,\vec{\sigma}\cdot(\vec{P}\times\vec{L}-\vec{L}\times\vec{P})
-\frac{i\mu}{r}\lambda P_{5}
\end{equation}
which lead to 
\begin{equation}
\{\alpha^{*},\lambda\}=\frac{2i}{\sqrt{V}}\,\lambda P_{5}\,,\quad 
\{\alpha,\lambda\}=-\frac{2i}{\sqrt{V}}\,\lambda P_{5}\,. 
\end{equation}

\end{document}